\documentclass[aps,prl,reprint,superscriptaddress]{revtex4-2}

\usepackage[dvipdfmx]{graphicx}
\usepackage[dvipdfmx,colorlinks=true,citecolor=blue,linkcolor=blue,urlcolor=blue]{hyperref}
\usepackage{notoccite}
\begin{document}


\title{Commensurate--Incommensurate Transition in Submonolayer $^3$He on Graphite}


\author{A. Kumashita}
\email[]{ri21x011@gmail.com}
\affiliation{Graduate School of Science, University of Hyogo, 3-2-1, Kouto, Kamigoricho, Ako-gun, Hyogo, 678-1297, Japan}
\author{J. Usami}
\email[]{j-usami@aist.go.jp}
\affiliation{National Institute of Advanced Industrial Science and Technology (AIST), Central 4-1, 1-1-1, Higashi, Tsukuba, Ibaraki, 305-8565, Japan}
\author{S. Komatsu}
\affiliation{Graduate School of Science, University of Hyogo, 3-2-1, Kouto, Kamigoricho, Ako-gun, Hyogo, 678-1297, Japan}
\author{Y. Yamane}
\affiliation{Graduate School of Science, University of Hyogo, 3-2-1, Kouto, Kamigoricho, Ako-gun, Hyogo, 678-1297, Japan}
\affiliation{Graduate School of Engineering Science, The University of Osaka, 1-3, Machikaneyamacho, Toyonaka, Osaka, 560-8531, Japan}
\author{S. Miyasaka}
\affiliation{Graduate School of Science, University of Hyogo, 3-2-1, Kouto, Kamigoricho, Ako-gun, Hyogo, 678-1297, Japan}
\author{H. Fukuyama}
\email[]{hiroshi@kelvin.phys.s.u-tokyo.ac.jp}
\affiliation{Cryogenic Research Center, The University of Tokyo, 2-11-16, Yayoi, Bunkyo-ku, Tokyo, 113-0032, Japan}
\author{A. Yamaguchi}
\email[]{yamagu@sci.u-hyogo.ac.jp}
\affiliation{Graduate School of Science, University of Hyogo, 3-2-1, Kouto, Kamigoricho, Ako-gun, Hyogo, 678-1297, Japan}


\date{\today}

\begin{abstract}
We report high-precision heat-capacity measurements of submonolayer $^3$He adsorbed on highly crystalline graphite, revealing new aspects of the commensurate--incommensurate transition. 
Below 1\,K, two possible striped domain-wall phases emerge: $\alpha_1$ with variable wall spacing and $\alpha_2$ with fixed spacing.
The $T$-linear heat capacity in $\alpha_1$ arises from one-dimensional phonons  along the walls.
$\alpha_2$ melts into $\alpha_1$ at a critical density via a second-order transition, consistent with a quantum nematic (quantum liquid-crystal) state in $\alpha_1$, and reconciling thermodynamic and prior nuclear-magnetic data.
\end{abstract}


\maketitle


The commensurate--incommensurate transition is a ubiquitous phenomenon in condensed matter systems, occurring, for example, in surfaces, magnets, and dielectrics.  
It gives rise to rich emergent behaviors, including discommensurations~\cite{P-Bak-D-Mukamel-J-Villain-and-K-Wentowska1979-lv, bak1982commensurate}, soliton excitations~\cite{Pokrovsky1980-ox, kartashov2011solitons}, charge-density waves~\cite{gruner1988dynamics}, and moir\'e superlattices~\cite{he2021moire}, highlighting the pivotal role of competing interactions, frustration, topological defects, and fluctuations in low-dimensional systems. 

\begin{figure}[b]
  \includegraphics[width=\columnwidth]{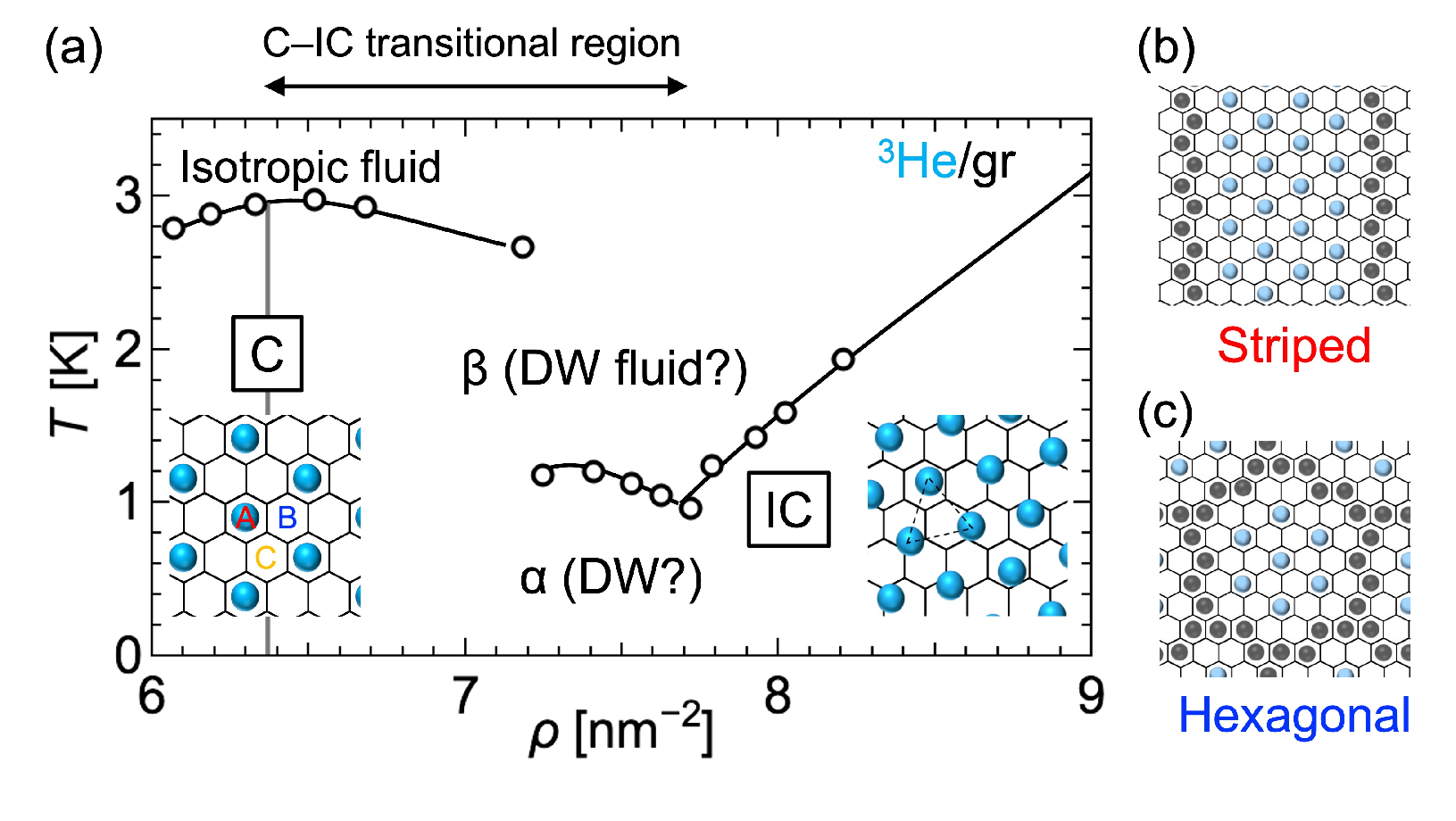}
  \caption{
(a) Phase diagram of submonolayer $^3$He on graphite inferred from previous heat-capacity measurements~\cite{hering1976apparent, Bretz1973-vs}. Open circles denote heat-capacity peak temperatures. 
Phases: C, $\sqrt{3} \times \sqrt{3}$ commensurate solid; $\alpha$, putative DW phase; $\beta$, DW fluid; IC, incommensurate solid.  
Insets: schematic structures of the C and IC solid phases; A, B, and C mark the three equivalent adsorption sites in the C phase. 
Unrelaxed (b) striped and (c) hexagonal arrangements of superheavy DWs. 
Black spheres represent atoms in the DWs, and light blue spheres denote those in the domains.
}
  \label{Fig_phase_predicted}
\end{figure}

Submonolayer films physisorbed on graphite (gr) provide a prototypical two-dimensional (2D) system in which adatom--adatom interactions compete with adatom--substrate interactions~\cite{dash1980phase, taub2012phase, Nijs1988phase}. 
At the stoichiometric density $\rho_{1/3} = 6.366~\mathrm{nm}^{-2}$, a $\sqrt{3} \times \sqrt{3}$ commensurate solid (the C phase) forms, with atoms occupying one of three hollow sites in the graphite honeycomb lattice (see Fig.~\ref{Fig_phase_predicted}(a)).  
At higher densities ($\rho \geq$ 7.6~nm$^{-2}$), a 2D triangular lattice solid incommensurate with the graphite substrate (the IC phase) is stabilized due to stronger particle-particle interactions.
%
%
In the relatively broad intermediate C--IC density regime, novel phases can emerge from the competing interactions, most notably domain-wall (DW) phases~\cite{hering1976apparent,halpin1986observation}.
For {\it classical} systems, it is well established 
that Xe/gr~\cite{Peters1989-xl, Grimm1999-sz} exhibits a hexagonal DW phase at $\rho < \rho_{1/3}$, 
whereas Kr/gr~\cite{Moncton1981-wv, coppersmith1982dislocations} hosts a thermally melted DW phase---a reentrant fluid---at $\rho > \rho_{1/3}$, 
known as the DW fluid.
In contrast, for {\it quantum} systems, H$_2$/gr and HD/gr are known to have only striped DW phases, whereas D$_2$/gr, a slightly less quantum system, undergoes transitions from striped-to-hexagonal DW phases with increasing density~\cite{freimuth1986specific, freimuth1987commensurate, cui1989low, wiechert1991ordering, wiechert1992heat, Wiechert2003physics}. 
With increasing temperature, the ordered DW phases melt into an isotropic fluid through the DW fluid ($\beta$) phase.
%
%
%
%
%
%
%
 
%
For $^{3}$He/gr, the {\it most quantum} submonolayer system, relatively little is known about its C--IC transition. 
High-temperature heat capacity measurements~\cite{hering1976apparent,Bretz1973-vs} suggest a phase diagram (Fig.~\ref{Fig_phase_predicted}(a)) that is qualitatively similar to those of H$_2$/gr and HD/gr.
However, the detailed DW structure, including both in $^3$He/gr and in $^4$He/gr, its bosonic counterpart, remains unresolved, as neutron-diffraction studies have detected no satellite peaks~\cite{lauter1987neutron, Lauter1991, godfrin1995experimental}. 
Theories, however, predict novel phenomena, such as a quantum phase transition from striped~(Fig.~\ref{Fig_phase_predicted}(b)) to hexagonal DW symmetry~(Fig.~\ref{Fig_phase_predicted}(c))~\cite{halpin1986observation,corboz2008phase} or  quantum melting of the striped DW phase~\cite{momoi2003quantum}---namely, emergence of a quantum DW fluid. 
%
%
%
%
%
At ultralow temperatures below 10~mK, nuclear-spin heat capacity~\cite{Greywall1990-xj} and nuclear magnetization~\cite{Siqueira1992-qp,Rapp1993-ld,Ikegami1998-cd} of $^3$He/gr
exhibit an intriguing density dependence
in the C--IC region.
%
Because nuclear-spin exchange interactions in a localized $^3$He system originate from atom-atom tunneling,
the magnetic properties provide valuable insights into the underlying quantum state.
Nevertheless, the relationship between the magnetic behavior and the DW symmetry remains unclear.
%
%
 
%
In this Letter, we report high-precision heat-capacity measurements of submonolayer $^3$He on graphite in the C--IC region using a ZYX graphite substrate, over the temperature range 0.3--3.5\,K.  
ZYX has a platelet size of 100--300\,nm~\cite{birgeneau1982high,niimi2006scanning}, providing microcrystallites approximately 100 times wider than those of Grafoil used in previous studies. 
The use of this higher crystalline substrate enables us to resolve subtle features that were previously masked.  
Our data reveal that a DW ($\alpha_{1}$) phase with variable wall density, most likely a striped DW phase, appears from an areal density remarkably close to the C phase, and it transforms into another DW ($\alpha_{2}$) phase of the same class with a fixed DW density above a critical density. 
Comparison with previous nuclear-magnetization data~\cite{Greywall1990-xj,Ikegami1998-cd} suggests that the $\alpha_{1}$--$\alpha_{2}$ transition corresponds to quantum melting (or freezing) of the DW structure, 
consistent with the theoretical prediction~\cite{momoi2003quantum}.
%
%
 
%
%
The experimental setup is described in Ref.~\cite{usami2021superconducting}.  
Heat-capacity measurements were performed using a quasi-adiabatic heat-pulse method.  
The adsorption surface area of the ZYX substrate is $A = 27.3 \pm 0.5\,\mathrm{m}^2$.  
The coverage scale is calibrated using both the C-phase density ($\rho_{1/3}$) and the second-layer promotion density (11.2\,nm$^{-2}$)~\cite{nakamura2016possible}, as described in Supplemental Material (SM)-1~\cite{SuppMat}.  
The addendum contribution was subtracted from the total measured heat capacity. 
The contribution was minimized by using a calorimeter made of superconducting niobium, which compensates for ZYX’s ten times smaller specific surface area than Grafoil~\cite{usami2021superconducting}.
%
%
%
%
%
%
 
\begin{figure}[t]
  \includegraphics[width=\columnwidth]{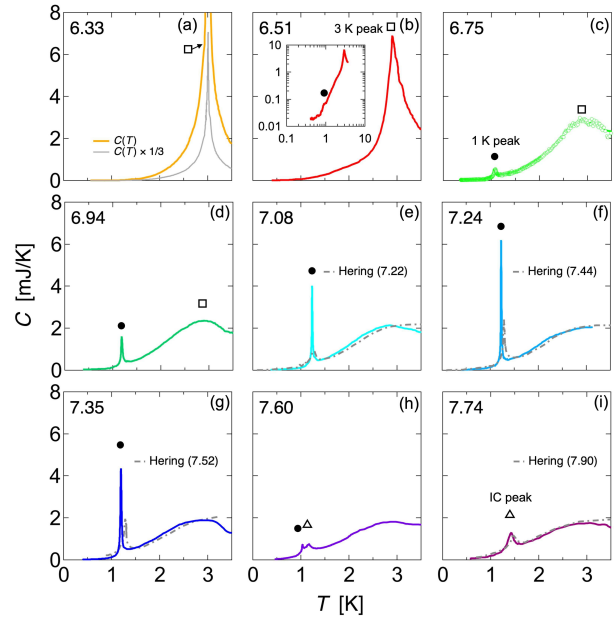}
  \caption{
Heat-capacity data for submonolayer $^3$He films on a ZYX graphite substrate at selected areal densities. 
Complete datasets for measured densities are provided in SM-2~\cite{SuppMat}.
Raw data are shown only for $\rho$ = 6.75~nm$^{-2}$; otherwise, the data are shown as smoothed curves for clarity.
Squares, circles, and triangles indicate the 3\,K, 1\,K, and IC peaks, respectively.
Dash-dotted curves reproduce data from Ref.~\cite{hering1976apparent} on Grafoil (see SM-3~\cite{SuppMat} for density rescaling). 
Numbers denote areal densities in nm$^{-2}$.
Inset: smallest 1\,K peak detected at $\rho_{\mathrm{1}} = 6.51\ \mathrm{nm}^{-2}$.  
}
  \label{Fig_HCdata}
\end{figure}

%
Figures~\ref{Fig_HCdata}(a)--(i) show heat-capacity data at several representative densities.
Three distinct anomalies are observed: the ``3\,K peak," ``1\,K peak," and ``IC peak," each associated with a different phase transition. 
(a) At $\rho = 6.33\ \mathrm{nm}^{-2}$ near $\rho_{1/3}$, a sharp heat-capacity peak with long high- and low-temperature tails appears around 3\,K (``3\,K peak," square symbol), which is associated with the order-disorder (second-order) transition of the C phase~\cite{Bretz1973-vs}.   
The high crystallinity of ZYX is evidenced by the peak height $C_{\mathrm{peak}} =$ 9.0$Nk_{\mathrm{B}}$, substantially larger than the 5.0$Nk_{\mathrm{B}}$ reported for Grafoil~\cite{Bretz1973-vs}.
Here, $N$ is the total number of $^3$He atoms, and $k_{\text{B}}$ is the Boltzmann constant.
(b) With slight excess density, the 3\,K peak rapidly broadens, and a small but sharp peak immediately appears near 1\,K (dot) above 6.51\,nm$^{-2}$ ($\rho_{\mathrm{1}}$), marking the onset of the ``1\,K peak," (inset).

(c)--(e) As the density increases further, the 1\,K peak progressively grows and sharpens, reaching its maximum at 7.19--7.25\,nm$^{-2}$ ($\rho_{\mathrm{2}}$) [panel (f)]. 
The maximum 1\,K peak height is 2.8 times larger than that for Grafoil~\cite{hering1976apparent} (dash-dotted line), highlighting a pronounced finite-size effect. 
Likewise, the peak is narrower than in $^4$He/Grafoil~\cite{greywall1993heat} by the same factor.
In the same density range, the 3\,K anomaly evolves into a broad bump, consistent with a crossover from a DW fluid to an isotropic fluid. 
(g) Above $\rho_{\mathrm{2}}$, the 1\,K peak rapidly diminishes up to 7.57\,nm$^{-2}$ ($\rho_{\mathrm{3}})$ and (h) disappears through a narrow two-phase coexistence region, where an additional broad peak (the ``IC peak," triangle) appears, producing a double-peak structure.
The density span of this two-phase region is only 0.03\,nm$^{-2}$.
(i) The IC peak is attributed to the 2D melting of the IC solid~\cite{hering1976apparent} and is therefore intrinsically broad, exhibiting little to no size effect.
%
%
%
 
%
%
%

%
\begin{figure}[t]
  \includegraphics[width=\columnwidth]{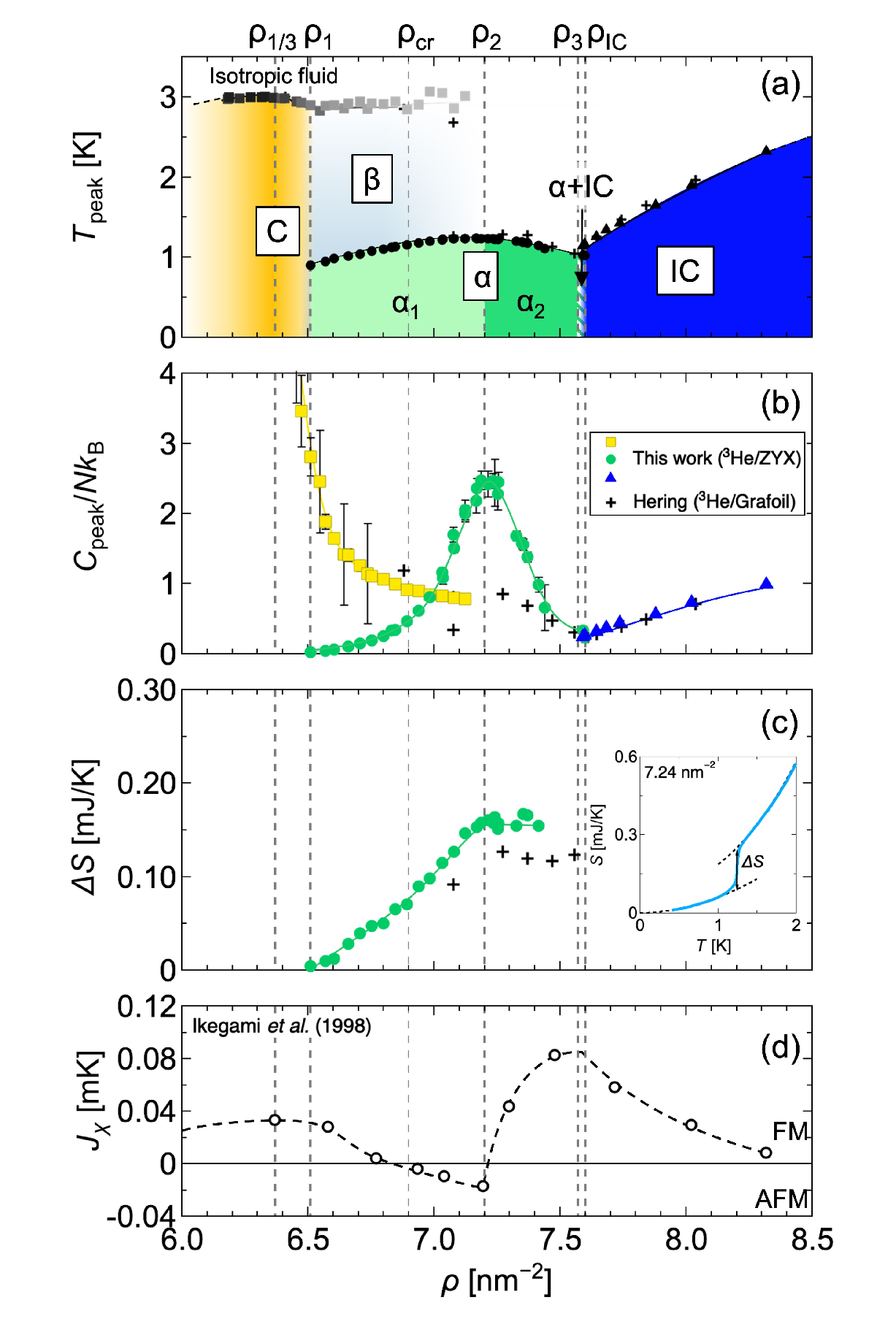}
  \caption{
Density dependences of heat-capacity anomalies: (a) peak temperature $T_{\mathrm{peak}}$; (b) specific-heat peak height $C_{\mathrm{peak}}/N k_{\mathrm{B}}$; (c) entropy change $\Delta S$ from the 1\,K peak. 
The proposed phase diagram of the submonolayer $^3$He on graphite is shown in color in (a); $\rho_{\mathrm{1/3}}$, $\rho_{1}$, $\rho_{2}$, $\rho_{3}$, $\rho_{\mathrm{IC}}$, and $\rho_{\text{cr}}$ denote characteristic areal densities (see text). 
In (b), yellow squares, green circles, and blue triangles represent the 3\,K, 1\,K, and IC peaks, respectively. 
In (a)--(c), crosses are data from Ref.~\cite{hering1976apparent} on Grafoil (see SM-3~\cite{SuppMat} for density rescaling). 
Inset: temperature dependence of the entropy $S$ deduced from the present heat-capacity data at $\rho = 7.24\ \mathrm{nm}^{-2}$. 
%
%
%
%
(d)~Effective exchange energy $J_{\chi}$ determined from the nuclear-magnetization measurement~\cite{Ikegami1998-cd} (see text). 
Solid and dashed lines are guides to the eye.
}
  \label{Fig_Cpeak_Tpeak}
\end{figure}
%
%
Figures~\ref{Fig_Cpeak_Tpeak}(a)--(c) summarize the density dependences of (a) the heat-capacity peak temperature $T_{\text{peak}}$, (b) specific-heat peak height $C_{\text{peak}}/Nk_{\text{B}}$, and (c) the entropy changes $\Delta S$ associated with the 1\,K heat-capacity anomalies, as determined from the present data. 
The high crystallinity of ZYX reveals fine details of the C--IC transition in this system. 
%
The resulting phase diagram is shown in color in Fig.~\ref{Fig_Cpeak_Tpeak}(a). 
Given the qualitative similarity to the phase diagrams of H$_2$/gr and HD/gr~\cite{freimuth1986specific, freimuth1987commensurate, cui1989low, wiechert1991ordering, wiechert1992heat, Wiechert2003physics} (see SM-4~\cite{SuppMat}), where neutron-scattering experiments have been performed, it is reasonable to associate the 1\,K peak in $^3$He/gr with melting of a striped DW phase (hereafter the $\alpha$ phase) to a DW fluid phase (hereafter the $\beta$ phase). 
%
%

%
%
%
%
%
%
%
Importantly, we observe clear discontinuities in the density dependence of $C_{\text{peak}}$ (Fig.~\ref{Fig_Cpeak_Tpeak}(b)) and $\Delta S$ (Fig.~\ref{Fig_Cpeak_Tpeak}(c)), i.e., in $dC_{\text{peak}}/d\rho$ and $d(\Delta S)/d\rho$, 
indicating a second-order phase transition within the striped DW phase at $\rho_{\mathrm{2}}$.
We define the $\alpha$ phase regions below and above $\rho_{\mathrm{2}}$ as the $\alpha_1$ and $\alpha_2$ phases, respectively.
To quantify the entropy jump, 
we first deduce the entropies from the measured heat capacity by numerically integrating $C(T)/T$, and evaluate $\Delta S$ as the difference between the extrapolated high- and low-temperature behaviors of $S(T)$ at $T_{\text{peak}}$ (dotted lines in the inset of Fig.~\ref{Fig_Cpeak_Tpeak}(c)) (SM-5~\cite{SuppMat}).
%
%
The distinct entropy jump, together with the sharp heat-capacity peak, indicates that the 1\,K anomaly corresponds to a first-order (melting) transition to the $\beta$ phase, as previously discussed for $^4$He/Grafoil~\cite{greywall1993heat}.
This assignment is further supported by a criticality analysis of the 1\,K peak: fitting the data to $C \propto t^{-\alpha}$, with $t = |T - T_{\mathrm{peak}}|/T_{\mathrm{peak}}$, yields a large exponent $\alpha$ = 0.8--1.3, implying that the transition is {\it not} second order (see SM-6~\cite{SuppMat}).

%
Remarkably, the 1\,K peak appears already very close to $\rho_{\mathrm{1/3}}$ with only a 2.3\% excess density and at $T_{\text{peak}} = 0.90$\,K. 
%
%
The data clearly indicate that the $\beta$ phase does not intervene between the C and $\alpha$ phases at finite temperatures (see SM-4~\cite{SuppMat}). 
This observation is consistent with the theoretical prediction~\cite{P-Bak-D-Mukamel-J-Villain-and-K-Wentowska1979-lv, bak1982commensurate, coppersmith1982dislocations, Pokrovsky1980-ox} for classical $p = 3$ systems, where $p$ denotes the order of the ground-state degeneracy (see bottom-left cartoon in Fig.~\ref{Fig_phase_predicted}(a)). 
To our knowledge, this is the first definitive experimental confirmation of this behavior.
In contrast, for $p = 2$ (Ising) systems, the theories predict that the $\alpha$ phase is unstable, resulting in the appearance of a $\beta$ phase in between the C and $\alpha$ phases at finite temperatures, as confirmed experimentally in D$_2$/Kr/gr~\cite{Wiechert2004-wv}.
Relatedly, we identified a narrow density window between $\rho_{\mathrm{1/3}}$ and $\rho_{\mathrm{1}}$, where the 1\,K peak has not yet emerged, although the rather sharp 3~K peak remains.
This region requires further investigation and may represent a feature unique to the quantum $p = 3$ system, which is not captured by existing classical theories.
%

%
Across the entire C--IC region, $T_{\text{peak}}$ varies by only 27\% and exhibits a broad maximum without singularities at $\rho_{\mathrm{2}}$.
This smooth, weak density dependence of $T_{\text{peak}}$ indicates that no structural phase transition---such as a striped-to-hexagonal DW transition---occurs at $T = 0$ as a function of the density, since such transitions are believed to be first order~\cite{P-Bak-D-Mukamel-J-Villain-and-K-Wentowska1979-lv,bak1982commensurate,halpin1986observation,coppersmith1982dislocations,freimuth1986specific, freimuth1987commensurate, cui1989low, wiechert1991ordering, wiechert1992heat, Wiechert2003physics}.
Instead, we observed clear discontinuities in $dC_{\text{peak}}/d\rho$ and $d(\Delta S)/d\rho$, indicating a second-order phase transition at $\rho_{\mathrm{2}}$ within the same $\alpha$-type phases ($\alpha_1$ and $\alpha_2$). 
%
%
%

%
Within the $\alpha_1$ phase, $\Delta S$ increases almost linearly with increasing density. 
If each DW releases a constant amount of entropy upon melting when DW--DW interactions are unimportant, 
then this linear increase in $\Delta S$ implies a linear increase in the number of DWs, i.e., the DW density. 
Therefore, assuming a striped structure composed of superheavy DWs (see Fig.~\ref{Fig_phase_predicted}(b)), the $\alpha_1$ phase is considered to have a variable-spacing striped DW structure.
Intriguingly, a similar structure has been proposed based on neutron-scattering experiments for H$_2$/gr and HD/gr~\cite{freimuth1987commensurate, wiechert1991ordering,wiechert1992heat,Wiechert2003physics}. 
Looking more closely at the data, the slope $d(\Delta S)/d\rho$ increases by 60\% above 6.90\,nm$^{-2}$ ($\rho_{\mathrm{cr}}$), indicating that DW--DW interactions become increasingly important above $\rho_{\mathrm{cr}}$.
%

%
Above $\rho_{\mathrm{2}}$, $\Delta S$ abruptly saturates at a nearly constant value ($\Delta S/Nk_{\text{B}} = 0.057$) in the $\alpha_2$ phase.
The density $\rho_{\mathrm{2}}$ corresponds, on our adopted density scale (SM-7~\cite{SuppMat}), to a commensurate configuration of linear DWs separated by six atomic rows, a six-row structure, (Fig.~\ref{Fig_phase_predicted}(b)).
No analogous phase transition to a commensurate DW configuration has been reported in H$_2$/gr and HD/gr, instead, the six-row structure is known to correspond to the maximally compressed DW density in those systems~\cite{freimuth1987commensurate, wiechert1991ordering, wiechert1992heat, Wiechert2003physics}.
Taken together, these results lead to important conclusions: 
In $^3$He/gr, there is a quantum second-order phase transition at $\rho_{\mathrm{2}}$ between the variables-spacing striped DW ($\alpha_1$) phase and the fixed-spacing striped DW ($\alpha_2$) phase. 
Furthermore, both phases thermally melt into the DW fluid ($\beta$) phase via the first-order transition around 1\,K.
%

\begin{figure}[t]
\centering
  \includegraphics[width=\columnwidth]{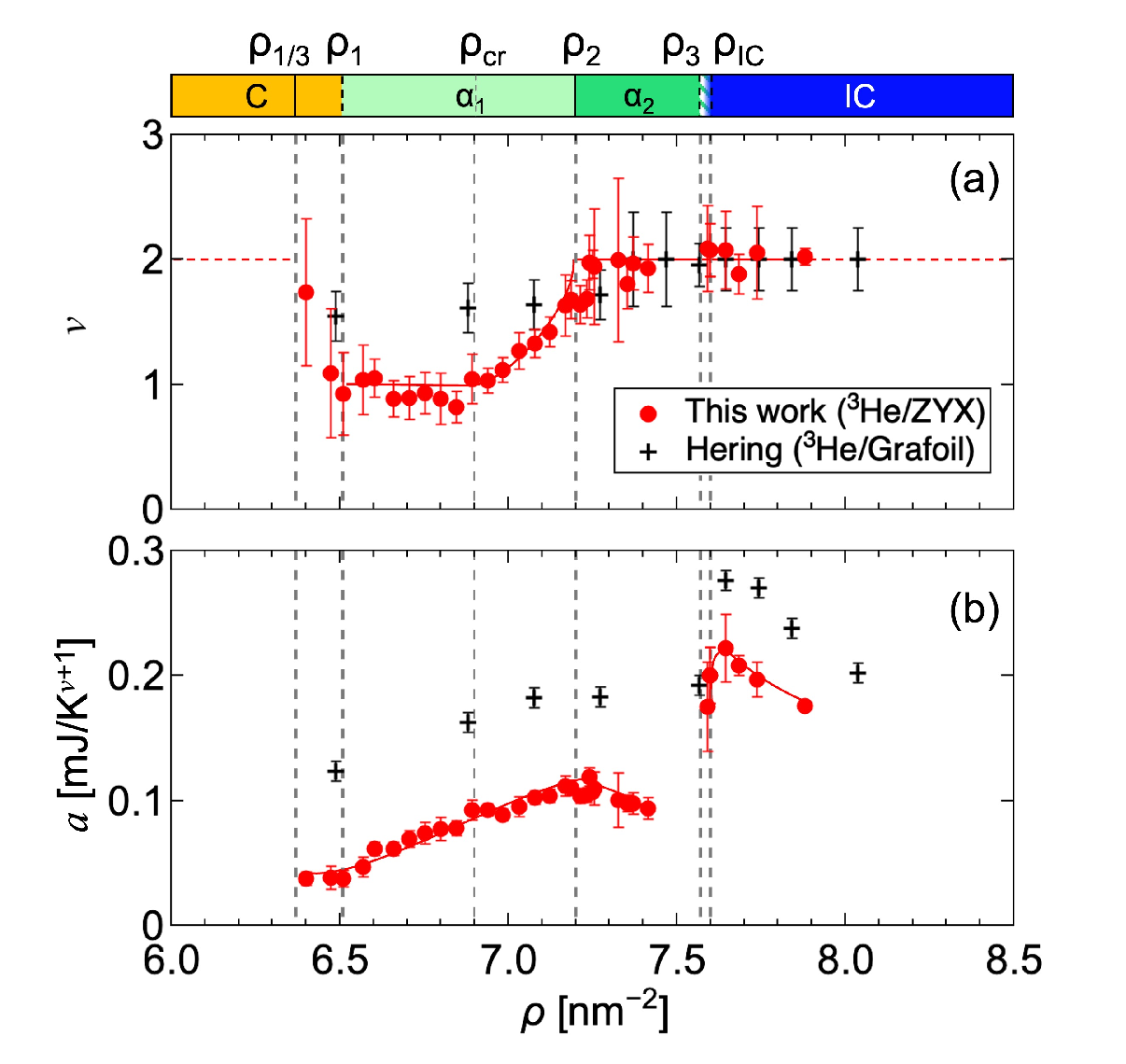}
  \caption{
Density dependence of (a) the exponent $\nu$ and (b) the coefficient $a$ obtained from power-law fits $C = a T^{\nu}$ over 0.3--0.9\,K.  
Red circles represent results from this study, while crosses show data from Ref.~\cite{hering1976apparent} on Grafoil (see SM-3~\cite{SuppMat}).  
Solid and dashed lines are guides to the eye.
The top panel shows the phase diagram proposed in this study.
Although Ref.~\cite{hering1976apparent} exhibits trends qualitatively similar to ours, its $a$ values are substantially larger, indicating strong finite-size effects in the phonon spectrum.
}
  \label{Fig_HC_exponent}
\end{figure}

%
%
Next, we focus on the low-temperature heat capacity below $T_{\text{peak}}$, investigating low-lying excitations in this system. 
Across a wide density range, the heat capacity follows a power-law, $C = a T^{\nu}$, between 0.3 and 0.9\,K~(see SM-2~\cite{SuppMat}). 
%
The fitted values of $\nu$ and $a$ are plotted in Figs.~\ref{Fig_HC_exponent}(a) and (b) as functions of $\rho$.
In the two solid phases (the C and IC phases), the exponent $\nu$ is approximately two, as expected for 2D phonons (Fig.~\ref{Fig_HC_exponent}(a)).
%
In contrast, in the C--IC region, $\nu$ evolves quite differently:
within the $\alpha_1$ phase, $\nu$ is close to unity up to the crossover density $\rho_{\text{cr}}$, and then gradually increases toward two at $\rho_{\mathrm{2}}$.
Meanwhile, the coefficient $a$ increases monotonically, independent of the crossover in $\nu$.
Across the density of $\rho_{\mathrm{2}}$, $\nu$ and $a$ exhibit discontinuous changes in their density dependence: $d\nu/d\rho$ and $da/d\rho$---$\nu$ becomes constant, whereas $a$ begins to decrease in the $\alpha_2$ phase---again, indicative of a second-order phase transition at $\rho_{\mathrm{2}}$.
%
 
%
The $T$-linear heat capacity ($\nu \approx 1$) below $\rho_{\text{cr}}$ in the $\alpha_1$ phase is intriguing, given that the system is a strictly 2D monolayer.
A preliminary heat-capacity measurement on $^4$He/gr using the same ZYX substrate also exhibits a $C \propto T$ dependence in the same density range, indicating an origin independent of quantum statistics (see SM-8~\cite{SuppMat}).  
Then the most likely origin is one-dimensional (1D) phonons, that are transverse phonons propagating along the linear DWs.
In this scenario, the low-temperature heat capacity is expressed as $C = (\pi^2/3) N_{\text{ex}} k_{\text{B}}(T/\theta_{\text{D}})$, where $N_{\text{ex}}$ and $\theta_{\text{D}}$ denote the number of excess He atoms relative to the C phase and the Debye temperature, respectively.
Applying this expression to the data for $\rho \leq \rho_{\text{cr}}$, we obtain $\theta_{\text{D}}$ increasing from 5~K to 8~K with increasing density.
%
%
%
%
%

Above $\rho_{\mathrm{cr}}$, repulsive DW interactions become significant, giving rise to a longitudinal mode propagating perpendicular to the DWs (anisotropic 2D phonon excitations). 
In the $\alpha_2$ phase, with fixed DW spacing above $\rho_{\mathrm{2}}$, increasing density stiffens the system (with $\theta_{\text{D}}$ progressively increasing and $a$ subsequently decreasing, assuming $N_{\text{ex}}$ remains nearly constant), consistent with an anisotropic 2D thermodynamic behavior. 
Upon freezing into the IC phase, an isotropic 2D solid emerges through a first-order transition between $\rho_{\mathrm{3}}$ and $\rho_{\mathrm{IC}}$, the transverse phonon mode propagating perpendicular to the ex-DWs appears, accounting for the abrupt increase in $a$. 
Subsequently, $a$ decreases due to increasing $\theta_{\text{D}}$~\cite{hering1976apparent}.
As described above, the low-temperature heat-capacities in the C--IC region are consistently explained by the phonon (Debye) model, further supporting the striped DW structure of the $\alpha_1$ and $\alpha_2$ phases.
%
%

%
We note a recent report of a similar crossover of $\nu$ from 1 to 2 within the C--IC region of a related system, namely $^4$He/gr with small amounts of $^3$He impurities~\cite{morishita2023fluidity}.
This behavior may correspond to the same phenomenon observed in our study. 
If so, it is best interpreted as a DW-interaction-driven crossover within a single striped DW phase as discussed here, rather than a striped-to-hexagonal DW structural transition, as suggested in Ref.~\cite{morishita2023fluidity}.
%

%
Finally, we consider the mechanism underlying the second-order quantum phase transition at $\rho_{\mathrm{2}}$.
A challenge for the $n$-row striped DW scenario, where $n$ is integer, in the $\alpha_1$ phase is the absence of the expected stepwise features at $n =7, 8, \ldots$, in any measured quantity. 
One argument invokes variation in $n$ arising from the finite platelet size and its distribution across the substrate~\cite{Wiechert2003physics,houston1970low,Schildberg1987-wh}.
However, this explanation is likely irrelevant for our $^3$He/gr data, because the ZYX platelet size is two orders of magnitude larger than the DW--DW separation ($l =$\,2.34 and 2.70\,nm for $n =$\,7 and 8, respectively (SM-7~\cite{SuppMat})), and the large zero-point motion should overcome DW pinning.
We therefore regard the observed smooth increase in DW density as intrinsic.

A comparison with previous nuclear-magnetization ($M$) data~\cite{Ikegami1998-cd} is instructive. 
As shown in Fig.~\ref{Fig_Cpeak_Tpeak}(d), the $M$ data exhibit an abrupt change in the density dependence of the exchange interaction ($J_{\chi}$), accompanied by a sign reversal at a density corresponding to $\rho_{\mathrm{2}}$ when density is rescaled by a factor of 0.99~\cite{SuppMat}. 
This indicates that the $\alpha_1$ phase is antiferromagnetic (AFM; $J_{\chi} < 0$), whereas $\alpha_2$ is ferromagnetic (FM; $J_{\chi} > 0$), with the two separated by a second-order transition.
However, within static DW configurations it is difficult to account for AFM behavior in $\alpha_1$, since both the C and IC constituent phases are known to be FM~\cite{Ikegami1998-cd}, implying that their striped-DW assembly should also be FM.

We therefore propose that the static fixed-spacing striped DW ($\alpha_2$) phase undergoes quantum melting at $\rho_{\text 2}$. 
Below this density, a quantum fluid with short-range stripe correlations---a quantum {\it nematic}, i.e., a quantum liquid crystal exhibiting uniaxial symmetry breaking---emerges in the $\alpha_1$ region. 
This hypothesis naturally explains the observed features of the $\alpha_1$ phase: 
the continuous variation of the DW density (or spacing) as an expectation value from the many-body nematic ground-state, and the AFM spin interactions manifesting as a gapless spin-liquid character induced by density fluctuations analogous to a similar state in the second layer of $^3$He~\cite{Ishida1997-gy, fukuyama2008nuclear, nakamura2016possible,Usami2026-nh}.
Moreover, such melting has been theoretically predicted for 2D quantum systems adsorbed on substrates~\cite{momoi2003quantum}, where the transition is driven by the proliferation of scale-free long DW dislocations and exhibits a critical behavior consistent with the 3D XY universality class. 
The most likely candidate for the high-temperature $\beta$ phase is then a DW fluid with short-range correlations of hexagonal DW arrangements and higher entropy than striped arrangements~\cite{coppersmith1982dislocations}.  

%
%
%

%
%
%
In summary, we performed high-precision heat-capacity measurements of submonolayer $^3$He on ZYX graphite. 
In the C--IC transitional density region, we conclude that two striped domain-wall (DW) phases emerge below 1\,K via first-order transitions: a variable-spacing striped DW ($\alpha_1$) phase and fixed-spacing striped DW ($\alpha_2$) phase, distinguished by their DW spacing.
The $T$-linear heat capacity in $\alpha_1$ is attributed to 1D phonons propagating along the DWs. 
With increasing density, it crosses over to $T^2$ behavior due to repulsive DW-DW interactions,  undergoes a second-order quantum transition to the $\alpha_2$ phase with a fixed DW spacing of (most likely) six atomic rows, and finally experiences a first-order transition to the IC phase (an isotropic 2D solid).
Combined with previous nuclear-magnetization data, $\alpha_{1}$ is most likely a quantum nematic phase.
The phase diagram we determined indicates the absence of an intervening $\beta$ phase between the C and $\alpha_{1}$ phases at finite temperatures, consistent with theoretical predictions for triply degenerate ground-state systems ($p = 3$) but in sharp contrast to $p = 2$ systems.
Our results motivate further theoretical investigation and direct structural probes~\cite{Yamaguchi2022,kumashita2023simulations,kumashita2025growth} to achieve a comprehensive understanding of quantum C--IC physics, including quantum nematicity in atomic systems.
\nocite{nakamura2013preliminary,Nakamura_unpublished,Nakamura2012-op,Sato2012-gv,Tejwani1980-zg}
%
%
 
%
We thank Tsutomu Momoi, Tomoki Minoguchi, Kazushi Aoyama, Akihiko Sumiyama and Hiroo Tajiri for their helpful discussions. This study was partially supported by JSPS KAKENHI (Grant No. 24K00826, 22H03883), JST SPRING, Japan (Grant No.  JPMJSP2175), and the Sasakawa Scientific Research Grant from The Japan Science Society.
\end{document}



\title{Supplemental Material: \\
Commensurate-Incommensurate Transition in Submonolayer $^3$He on Graphite}


\author{A. Kumashita}
\email[]{ri21x011@gmail.com}
\affiliation{Graduate School of Science, University of Hyogo, 3-2-1, Kouto, Kamigoricho, Ako-gun, Hyogo, 678-1297, Japan}
\author{J. Usami}
\email[]{j-usami@aist.go.jp}
\affiliation{National Institute of Advanced Industrial Science and Technology (AIST), Central 4-1, 1-1-1, Higashi, Tsukuba, Ibaraki, 305-8565, Japan}
\author{S. Komatsu}
\affiliation{Graduate School of Science, University of Hyogo, 3-2-1, Kouto, Kamigoricho, Ako-gun, Hyogo, 678-1297, Japan}
\author{Y. Yamane}
\affiliation{Graduate School of Science, University of Hyogo, 3-2-1, Kouto, Kamigoricho, Ako-gun, Hyogo, 678-1297, Japan}
\affiliation{Graduate School of Engineering Science, The University of Osaka, 1-3, Machikaneyamacho, Toyonaka, Osaka, 560-8531, Japan}
\author{S. Miyasaka}
\affiliation{Graduate School of Science, University of Hyogo, 3-2-1, Kouto, Kamigoricho, Ako-gun, Hyogo, 678-1297, Japan}
\author{H. Fukuyama}
\email[]{hiroshi@kelvin.phys.s.u-tokyo.ac.jp}
\affiliation{Cryogenic Research Center, The University of Tokyo, 2-11-16, Yayoi, Bunkyo-ku, Tokyo, 113-0032, Japan}
\author{A. Yamaguchi}
\email[]{yamagu@sci.u-hyogo.ac.jp}
\affiliation{Graduate School of Science, University of Hyogo, 3-2-1, Kouto, Kamigoricho, Ako-gun, Hyogo, 678-1297, Japan}


\date{\today}


\maketitle
\tableofcontents
\clearpage

\section{Density scale calibration}
\label{sec:definition}
%
\nocite{P-Bak-D-Mukamel-J-Villain-and-K-Wentowska1979-lv,bak1982commensurate,Pokrovsky1980-ox,kartashov2011solitons,gruner1988dynamics,he2021moire,dash1980phase,taub2012phase,Nijs1988phase,hering1976apparent}
\nocite{halpin1986observation,Peters1989-xl,Grimm1999-sz,Moncton1981-wv,coppersmith1982dislocations,freimuth1986specific,freimuth1987commensurate,cui1989low,wiechert1991ordering,wiechert1992heat,Wiechert2003physics}
\nocite{bretz1977ordered,lauter1987neutron,Lauter1991,godfrin1995experimental,corboz2008phase,momoi2003quantum,Greywall1990-xj,Siqueira1992-qp,Rapp1993-ld,Ikegami1998-cd}
\nocite{birgeneau1982high,niimi2006scanning,usami2021superconducting,nakamura2016possible,SuppMat}
\nocite{greywall1993heat,Wiechert2004-wv,morishita2023fluidity,houston1970low,Schildberg1987-wh,Ishida1997-gy,fukuyama2008nuclear,Usami2026-nh,Yamaguchi2022,kumashita2023simulations,kumashita2025growth}

\begin{figure}[h]
  \includegraphics[width=\columnwidth]{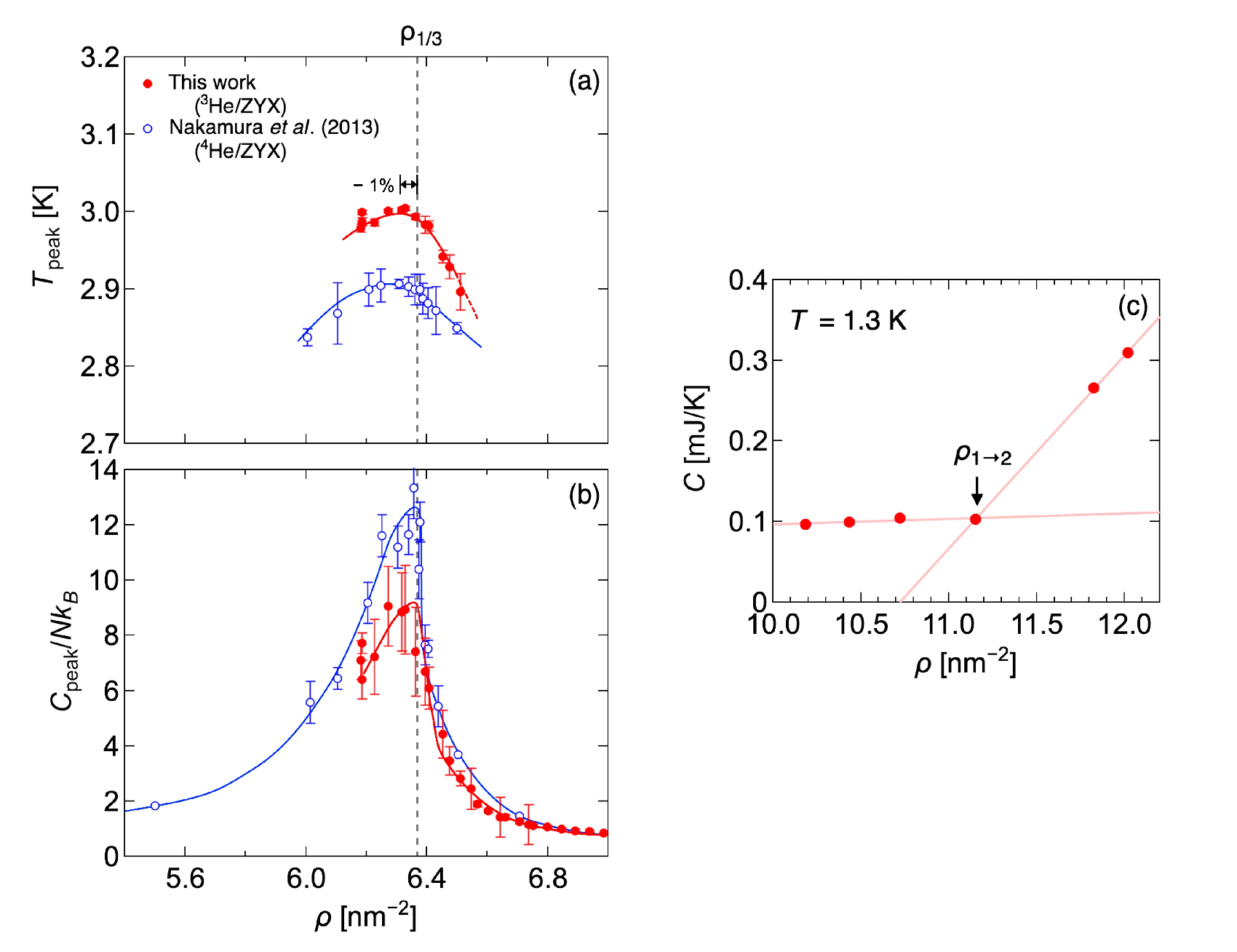}
  \caption{
Density dependences of (a) the 3~K peak temperature $T_\mathrm{peak}$ and (b) the specific-heat peak height $C_\mathrm{peak}/Nk_{\text{B}}$ for $^3$He/ZYX (this work; red closed circles) and $^4$He/ZYX~(Ref.~\cite{nakamura2013preliminary}; blue open circles). 
Vertical dashed line: stoichiometric density $\rho_{1/3}$ of the $\sqrt{3} \!\times\! \sqrt{3}$ commensurate (C) solid.
(c) Heat-capacity isotherms at 1.3~K near the second-layer promotion density $\rho_{\mathrm{1\rightarrow2}}$ (arrow) obtained in this work for $^3$He/ZYX. 
}
  \label{Fig_suppl_density_scale}
\end{figure}

The areal-density scale was calibrated using 
the stoichiometric density of the $\sqrt{3} \!\times\! \sqrt{3}$ commensurate (C) solid, $\rho_{1/3} = 6.366~\mathrm{nm^{-2}}$, at which both the specific-heat peak height $C_\mathrm{peak}/Nk_{\text{B}}$ and the 3\,K peak temperature $T_\mathrm{peak}$ are maximized, following earlier experiments on Grafoil~\cite{greywall1993heat,bretz1977ordered}.
More precisely, Nakamura {\it et al}.~\cite{nakamura2013preliminary}, who used a higher-crystallinity ZYX substrate, report that the $C_\mathrm{peak}/Nk_{\text{B}}$--$\rho$ peak is much sharper and asymmetric (about $T_\mathrm{peak}$) compared with Grafoil data, and that the density at which $T_\mathrm{peak}$ is maximized, $\rho_\mathrm{Tmax}$, is 1.0\% lower than at which $C_\mathrm{peak}/Nk_{\text{B}}$ is maximized, $\rho_\mathrm{Cmax}$, for $^4$He/ZYX. 
As shown in Figs.~\ref{Fig_suppl_density_scale}(a) and (b), although we did not measure this density region as precisely as in the 1\,K peak region, our $^3$He/ZYX data exhibit a similar offset between $\rho_\mathrm{Tmax}$ and $\rho_\mathrm{Cmax}$. 
Therefore, assuming the same 1.0\% shift between $\rho_\mathrm{Tmax}$ and $\rho_\mathrm{Cmax}$ for $^3$He/ZYX---an assumption verified for the same batch of ZYX used here~\cite{Nakamura_unpublished}---we evaluate the surface area of our ZYX substrate as $A = 27.3 \pm 0.5~\mathrm{m^2}$.
According to Ref.~\cite{nakamura2013preliminary}, this method yields a surface area consistent with that obtained from N$_{2}$ adsorption isotherms at $T$ = 77~K~\cite{Nakamura2012-op} and thus provides a reliable basis for density determination.
%
\par
%
%
The second-layer promotion density $\rho_{\mathrm{1\rightarrow2}}$ also serves as a reference point for density-scale calibration, particularly in the C--IC density region bounded by $\rho_{1/3}$ and $\rho_{\mathrm{1\rightarrow2}}$. 
When the areal density exceeds $\rho_{\mathrm{1\rightarrow2}}$, 
liquid puddles with a constant density (0.6--0.8\,nm$^{-2}$) form in the second layer~\cite{Sato2012-gv}, 
resulting in a linear increase in the total heat capacity at fixed temperature. 
Figure~\ref{Fig_suppl_density_scale}(c) shows the heat-capacity isotherm of $^3$He/ZYX at 1.3~K for $\rho=10$--12\,nm$^{-2}$. 
From the intersection of extrapolated low- and high-density trends, we determine $\rho_{\mathrm{1\rightarrow2}} = 11.16 \pm 0.22$~nm$^{-2}$. 
This value agrees, within experimental uncertainty, with the previously reported value of 11.2 $\pm$ 0.2~nm$^{-2}$ for $^3$He/ZYX~\cite{nakamura2016possible}.

In summary, the absolute uncertainty of our areal-density scale in the C--IC region is $\pm2$\%, whereas the relative uncertainty is smaller by about a factor of twenty. 

%
%
%
%
\clearpage
\section{Heat capacity data}
\begin{figure}[h]
  \includegraphics[width=0.9\columnwidth]{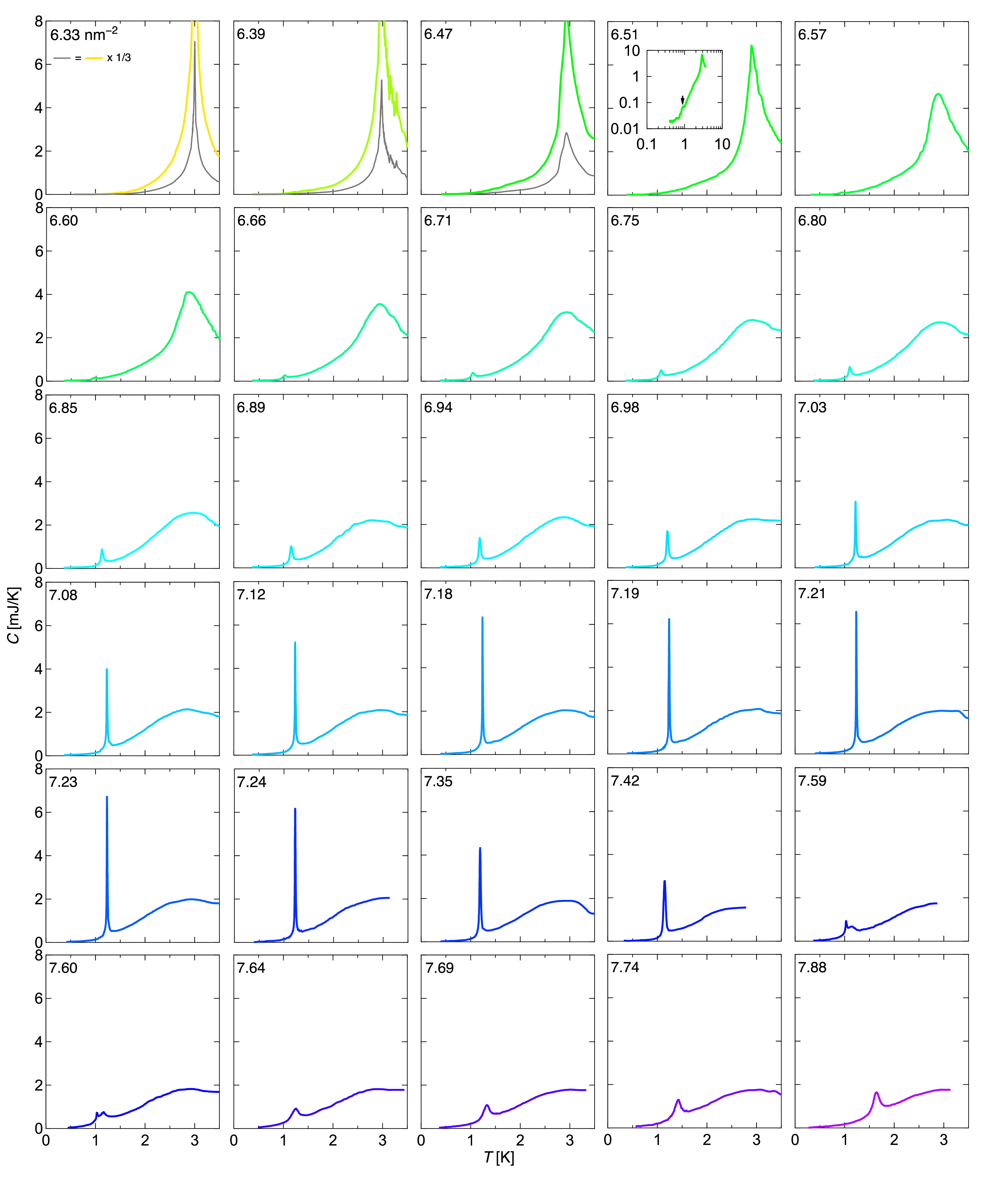}
  \caption{
Complete set of heat-capacity data for $^3$He/ZYX measured in this study. The number in the upper-left corner of each panel indicates the areal density in units of nm$^{-2}$.
}
  \label{Fig_suppl_HCdata}
\end{figure}

\begin{figure}[h]
  \includegraphics[width=0.8\columnwidth]{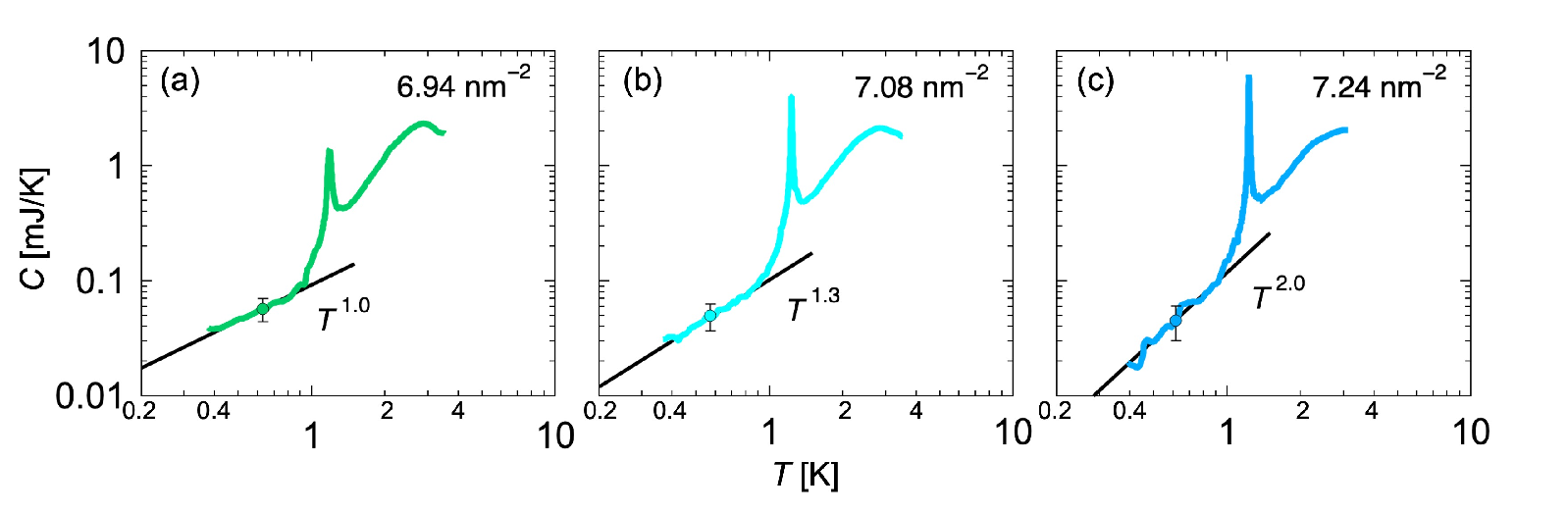}
  \caption{
(a--c) Log--log plots of the heat capacity at representative densities (6.94, 7.08, and 7.24\,nm$^{-2}$). 
Solid lines show fits to $C = a T^{\nu}$.
}
\vspace{15pt}
  \label{Fig_suppl_lowTHC}
\end{figure}

Figure~\ref{Fig_suppl_HCdata} compiles heat capacity data of $^3$He/ZYX obtained in this study. Log--log plots at representative densities (6.94, 7.08, and 7.24\,nm$^{-2}$) are shown in Fig.~\ref{Fig_suppl_lowTHC}. 
%
%
%
%
%
\section{Density rescaling of prior submonolayer studies using Grafoil}
\label{subsec:coverage}
%
%
\begin{figure}[b]
  \includegraphics[width=0.6\columnwidth]{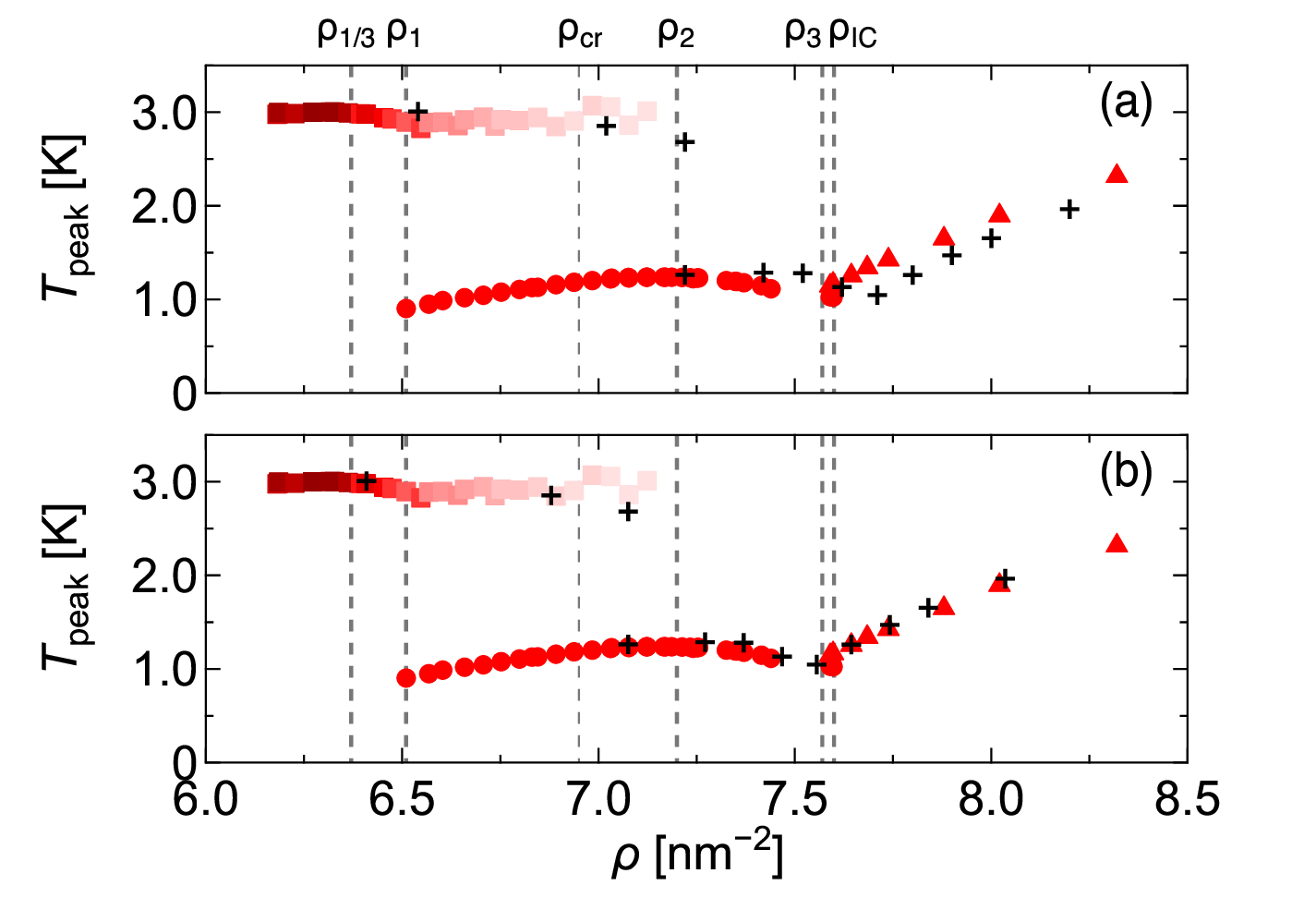}
  \caption{
Heat-capacity peak temperature $T_{\mathrm{peak}}$ as a function of areal density $\rho$ obtained for $^3$He/ZYX (this work; red symbols), compared with $^3$He/Grafoil data (Ref.~\cite{hering1976apparent}; crosses). (a) Original datasets without density rescaling.  
(b) After rescaling the $^3$He/Grafoil density by a factor $f = 0.980$ (see text).  
Labels $\rho_{\mathrm{1/3}}$, $\rho_{\mathrm{1}}$, $\rho_{\mathrm{2}}$, $\rho_{\mathrm{3}}$, $\rho_{\text{cr}}$, and $\rho_{\text{IC}}$ denote characteristic areal densities (see main text).  
%
%
}
  \label{Fig_suppl_Tpeak}
\end{figure}
%
Quantitative comparison with earlier work that used different density calibrations or different exfoliated graphite substrates (e.g., Grafoil) generally requires a density-scale correction of order of 2--8\%, as discussed extensively in Ref.~\cite{Usami2026-nh}.
This arises because exfoliated graphites contain certain surface heterogeneities ($\leq15$\%) associated with their platelet (microcrystallite) structure.

Figure~\ref{Fig_suppl_Tpeak}(a) compares our $T_{\mathrm{peak}}$ data with the Grafoil results of Hering \textit{et al}.~\cite{hering1976apparent}.  
Their $T_{\mathrm{peak}}$($\rho$) values (crosses) are shifted to higher densities relative to our $^3$He/ZYX data.  
However, upon rescaling their densities by a constant factor $f=0.980$, they agree very well with our data, as shown in Fig.~\ref{Fig_suppl_Tpeak}(b).
%
Thus the two density scales differ by 2\%, well within typical density-scale uncertainties noted above.
Moreover, a systematic offset may depend on substrate quality, at least in the submonolayer regime above $\rho_{\mathrm{1/3}}$. 
Given the higher crystallinity of ZYX compared with Grafoil and the $\sim$5\% Curie-constant deficit reported for $^3$He/Grafoil in nuclear magnetic susceptibility measurements~\cite{Rapp1993-ld,Ikegami1998-cd}, we therefore choose, for consistency in comparison, to use our density scale as a reference and to rescale the data of Hering \textit{et al}.~\cite{hering1976apparent} accordingly.

The validity of using a constant density-rescaling factor for Grafoil data is further supported by the agreement in the $T_{\mathrm{peak}}$--$\rho$ relation between our data and the $^4$He/Grafoil results of Greywall~\cite{greywall1993heat} after applying the same factor $f=0.980$ (not shown). \\  
%
%
%
\section{Role of ground state degeneracy in the C--IC transition}
\label{subsec:hydrogen}
\begin{figure}[b]
  \includegraphics[width=\columnwidth]{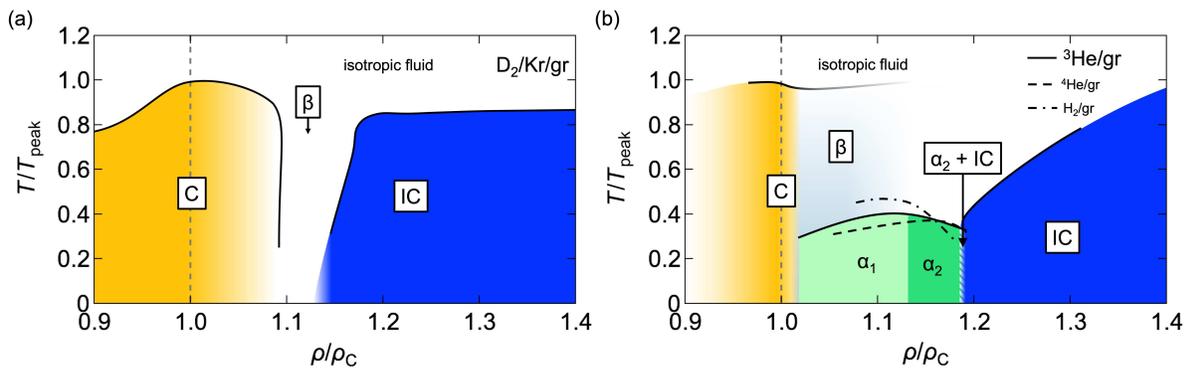}
  \caption{
Schematic phase diagrams of (a) D$_2$/Kr/gr~\cite{Wiechert2004-wv} and (b) $^3$He/gr (this work). 
%
Note that we do not represent the complex internal fine structure within the C and IC regions of D$_2$/Kr/gr inferred in Ref.~\cite{Wiechert2004-wv}.
In panel (b), the $\alpha$-$\beta$ transition lines for $^4$He/gr~\cite{greywall1993heat} and H$_2$/gr~\cite{freimuth1987commensurate} are also shown as the dashed and dash-dotted curves, respectively.
Phases: C, commensurate solid; $\alpha_1$, variable-spacing striped DW phase (nematic); $\alpha_2$, fixed-spacing striped DW phase; $\beta$, DW fluid; IC, incommensurate solid.
$\rho_{\text{C}}$ denotes the stoichiometric densities of the commensurate (C) phases.
}
  \label{Fig_suppl_hydrogen}
\end{figure}

Submonolayer $^{3}$He on graphite belongs to triply degenerate ground-state systems ($p = 3$; three-state Potts class), whereas D$_2$ on graphite preplated by krypton (D$_2$/Kr/gr) belongs to the $p = 2$ (Ising) class~\cite{Wiechert2004-wv}. 
Figure~\ref{Fig_suppl_hydrogen}(a) shows a schematic phase diagram of the D$_2$/Kr/gr system~\cite{Wiechert2004-wv}. 
The phase boundaries are plotted versus reduced density and temperature normalized to the melting (order-disorder) point of the commensurate (C) phase.
These phase lines are qualitatively distinct from those of $^{3}$He/gr ($p = 3$) in Figs.~3(a) and \ref{Fig_suppl_hydrogen}(b). 
In D$_2$/Kr/gr, a DW fluid ($\beta$) phase intervenes between the C and IC phases and is expected to remain stable down to very low temperatures~\cite{Wiechert2004-wv}.
By contrast, in $^{3}$He/gr, our heat capacity data reveal striped DW phases ($\alpha_{1}$ and $\alpha_{2}$) between C and IC, with $\alpha_{1}$ stabilizing extremely close to $\rho_{1/3}$ (see main text), indicating no intervening fluid phase there.
This provides the first definitive confirmation that no intervening $\beta$ phase exists in $p = 3$ systems, since the $\alpha\text{--}\beta$ transition lines reported for $^4$He/gr~\cite{greywall1993heat} and H$_2$/gr~\cite{freimuth1987commensurate} become rapidly ill-defined as $\rho$ approaches $\rho_{\text{C}}$ (see Fig.~\ref{Fig_suppl_hydrogen}(b)).
%
%

Such a qualitatively significant difference in the C--IC phase diagrams between $p = 2$ and $p = 3$ systems is consistent with theoretical predictions~\cite{bak1982commensurate,coppersmith1982dislocations}: when $p > \sqrt 8$, bound dislocation pairs play a crucial role in stabilizing striped DW phases, according to the Kosterlitz-Thouless criterion.
Having elucidated the C--IC transition in a quantum $p = 3$ system ($^{3}$He/gr), future studies on quantum $p = 2$ systems, such as He/Kr/gr~\cite{Tejwani1980-zg}, will be of great interest.

%
%
\section{Determination of entropy change $\Delta S$}
\label{sec:entropy}
\begin{figure}[h]
  \includegraphics[width=\columnwidth]{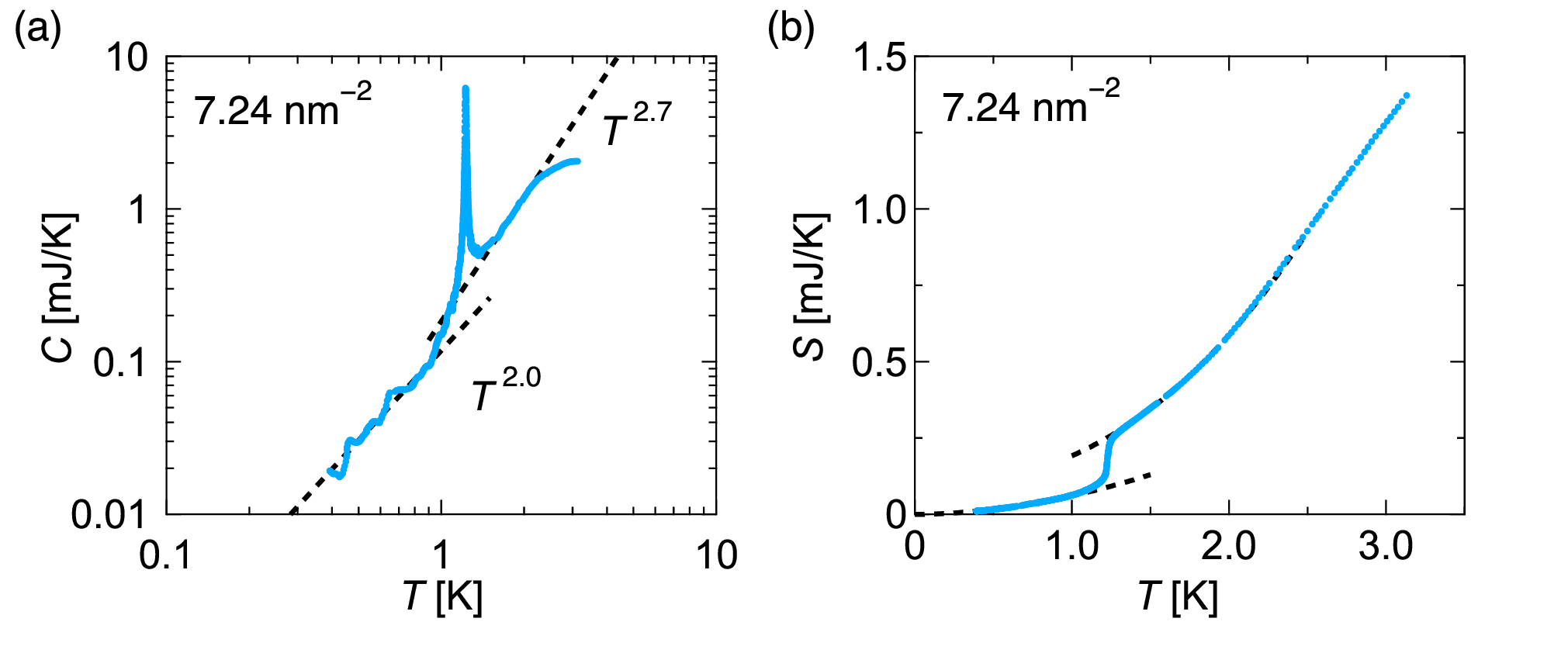}
  \caption{
(a) Log--log plot of the heat-capacity data at $\rho = 7.24$\,nm$^{-2}$.  
Dashed lines: power-law fits $C = a T^{\nu}$ on the low- and high-temperature sides of $T_{\mathrm{peak}}$. 
(b) Entropy $S(T)$ deduced from the dataset in (a).  
%
%
Dashed lines indicate $S = (a/\nu)T^{\nu}$ from the power-law fits.
}
  \label{Fig_suppl_entropy}
\end{figure}

Figure~\ref{Fig_suppl_entropy}(a) shows a log-log plot of the heat-capacity data at $\rho = 7.24$\,nm$^{-2}$ (same dataset as in Fig.~\ref{Fig_suppl_lowTHC}(c)).
Dashed lines represent power-law fits $C = a T^{\nu}$ to data on the low- and high-temperature sides of the peak temperature $T_{\mathrm{peak}}$, excluding the data in the critical or inhomogeneity-broadened region around $T_{\mathrm{peak}}$.
The fitting ranges are (0.42--0.78)$T_{\mathrm{peak}}$ and (1.22--1.94)$T_{\mathrm{peak}}$ for the low- and high-$T$ sides, respectively.

Figure~\ref{Fig_suppl_entropy}(b) shows the entropy $S(T)$ obtained by numerically integrating the measured dataset $\{(C_{i}$, $T_{i})\}$ as $S (T_{j}) = \sum_{i=1}^{j}(C_{i}/T_{i}) \Delta T_{i}$, while assuming the above power-law form for $C(T)$ below the lowest measured temperature $T_{1}$.
The expanded view near $T_{\mathrm{peak}}$ is shown in the inset of Fig.~3(c).  
Dashed curves plot $S = (a/\nu)T^{\nu}$, derived from the power-law fits to the $C(T)$ data.  
The entropy change $\Delta S$ is then defined as the difference, at $T_{\mathrm{peak}}$, between these extrapolated curves. 
%
%
%
%
%
\section{Criticality analysis of the 1\,K heat-capacity peak}
\label{sec:critical}
We analyze the nature of the phase transition (first vs second order) associated with the 1\,K heat-capacity peak, focusing on the sharpest peak at $\rho = 7.24$\,nm$^{-2}$.
To isolate the anomalous contribution---which may originate from either critical behaviors in a second-order transition or from inhomogeneity broadening in a first-order transition---we first determine the background heat-capacity $C_{\mathrm{BG}} (T)$.
As described in Sec.~\ref{sec:entropy} (Fig.~\ref{Fig_suppl_entropy}(a)), $C_{\mathrm{BG}}^{+}$ and $C_{\mathrm{BG}}^{-}$ are obtained by fitting power-law forms to the $C(T)$ data above and below $T_{\mathrm{peak}}$, respectively, and then subtracting them from the raw data.
The practical difficulty to draw a single smooth background curve across $T_{\mathrm{peak}}$ already points to first-order character, but we proceed with a quantitative critical analysis.

\begin{figure}[h]
  \includegraphics[width=\columnwidth]{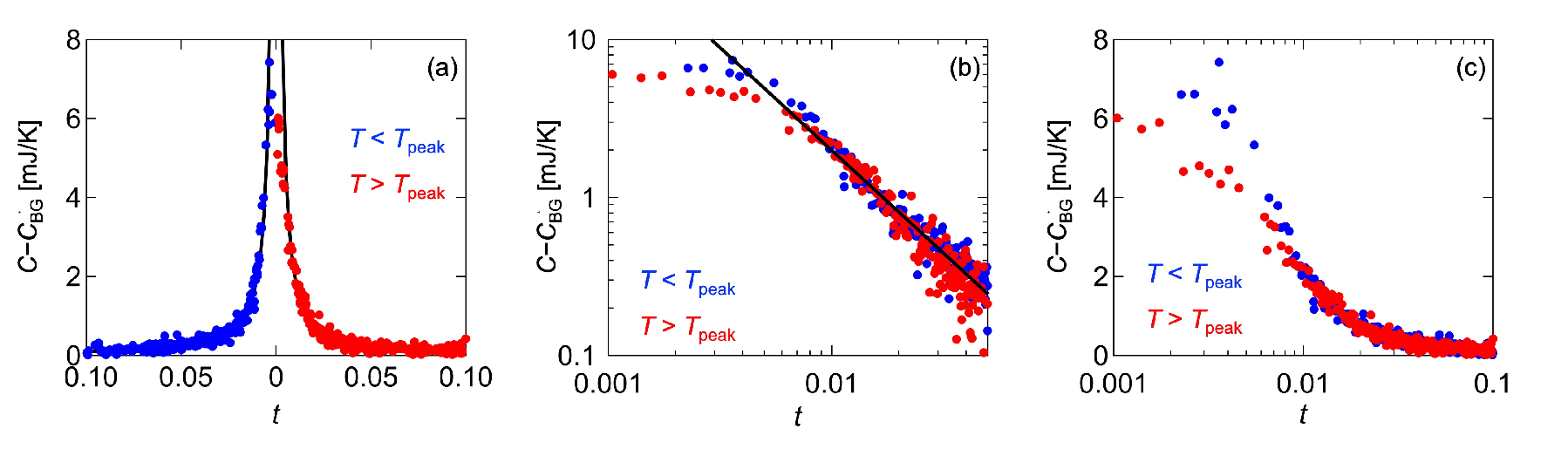}
  \caption{
Criticality analysis for the heat-capacity 1\,K peak of $^3$He/ZYX at $\rho = 7.24\ \mathrm{nm}^{-2}$.  
(a) Heat capacity after subtracting the background contribution $C_{\mathrm{BG}}^{\pm}$.  
(b) Log-log plot of the same dataset in (a) as a function of reduced temperature $t = |T - T_{\mathrm{peak}}|/T_{\mathrm{peak}}$.  
The solid line represents the fit to $C - C_{\mathrm{BG}}^{\pm} = D t^{-\alpha}$. (c) Semi-log plot of the same dataset in (a) and (b).
}
  \label{Fig_suppl_critical}
\end{figure}

The resulting anomalous parts, $C-C_{\mathrm{BG}}^{+}$ and $C-C_{\mathrm{BG}}^{-}$, are plotted in Figs.~\ref{Fig_suppl_critical}(a) and  \ref{Fig_suppl_critical}(b), as functions of the reduced temperature $t \equiv |T - T_{\mathrm{peak}}|/T_{\mathrm{peak}}$.  
The solid lines in Fig.~\ref{Fig_suppl_critical}(b) represent fits to
\renewcommand{\theequation}{S\arabic{equation}}
\begin{eqnarray}
C - C_{\mathrm{BG}}^{\pm} = D t^{-\alpha}, 
\label{eq-critical}
\end{eqnarray}
where $D$ is a coefficient and $\alpha$ the critical exponent; for this density we obtain $\alpha = 1.3\pm0.1$.  
Similar analyses at other densities yield $\alpha = 0.8\text{--}1.3$, which are too large for known second-order universality classes, favoring a first-order transition.
We note that the higher $t$ cutoff used to define the background (nonanomalous) fit ranges is $t = 0.2$--0.3, nearly independent of density.

Figure~\ref{Fig_suppl_critical}(c) shows a semi-log plot of $C - C^{\pm}_{\mathrm{BG}}$, which does not follow the linear behavior expected from a logarithmic divergence of the heat capacity in Ising-type systems; the same holds for other densities.
%
%
\section{Structures of the domain wall (DW) phases in $^3$He/gr}
\label{sec:DWspace}
\begin{figure}[h]
  \includegraphics[width=0.80\columnwidth]{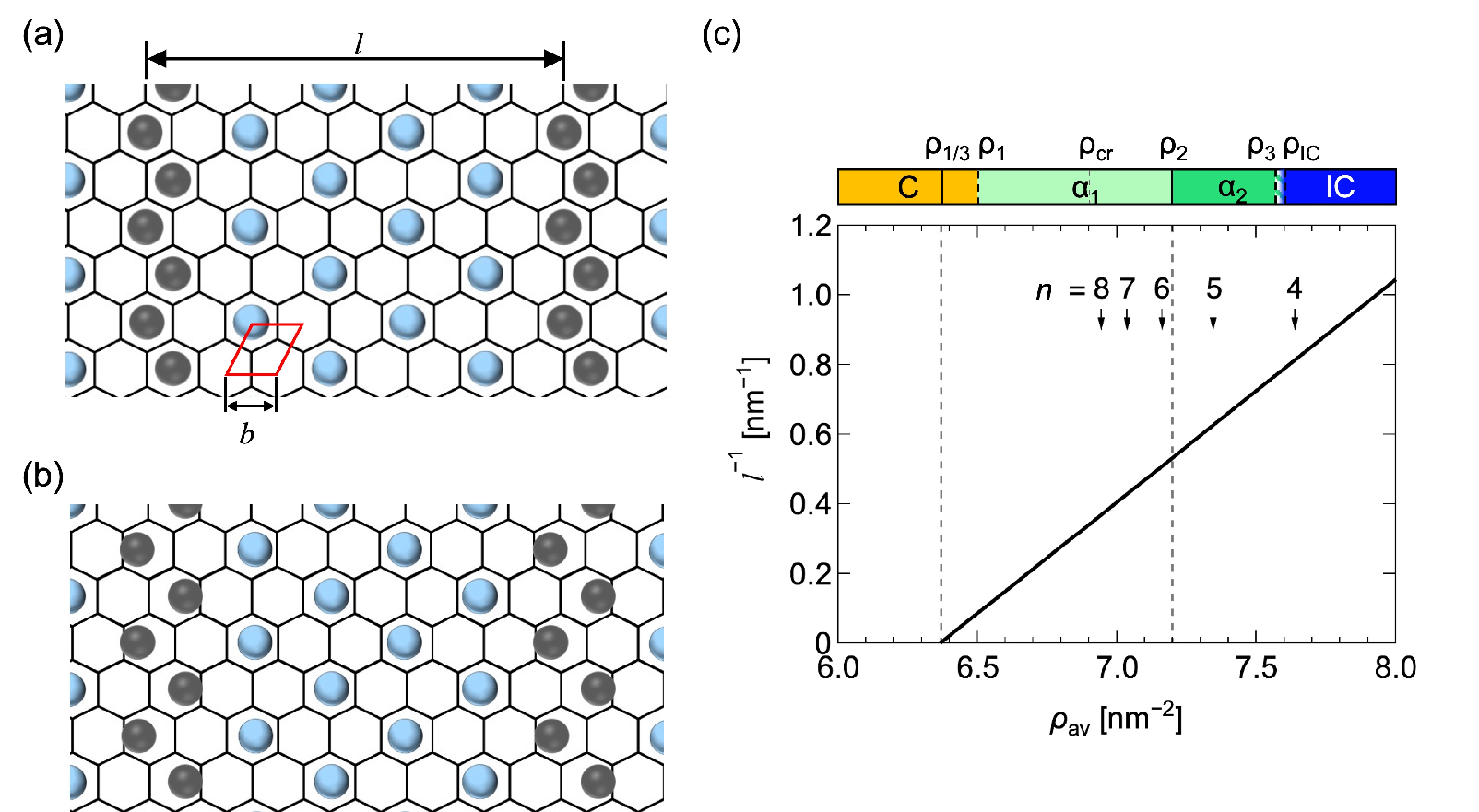}
  \caption{
(a) Schematic of the unrelaxed six-row superheavy striped DW structure. 
$l$ and $b$ denote the spacing between neighboring DWs and the graphite lattice constant, respectively.
(b) Relaxed six-row striped DW structure.
(c) Lower panel: inverse DW spacing $l^{-1}$ versus average areal density $\rho_{\text{av}}$ calculated from Eq.~\ref{eq-DWspacing3}; 
numbers mark densities corresponding to $n$(integer-$q$)-row striped DW structures.
Upper panel: phase diagram of the C--IC region in $^3$He/gr determined in this study.
}
  \label{Fig_suppl_DWdistance}
\end{figure}

Figure~\ref{Fig_suppl_DWdistance}(a) illustrates the six-row striped DW phase composed of {\it unrelaxed} superheavy walls, in which an additional atomic row is inserted every six rows into the underlying $\sqrt{3} \!\times\! \sqrt{3}$ commensurate (C) lattice.
In reality, owing to repulsive interatomic interactions arising from zero-point motion, the arrangement relaxes as shown in Fig.~\ref{Fig_suppl_DWdistance}(b)~\cite{Wiechert2003physics}, which is the most plausible structure for the $\alpha_{2}$ phase.

In general, the DW spacing $l$ in a striped DW phase with an effective (possibly noninteger) row number $q ( > 0)$ is 
\begin{eqnarray}
l = b(3q/2 - 1),
\label{eq-DWspacing1}
\end{eqnarray}
where $b= 0.2459$\,nm is the graphite honeycomb lattice constant (see the lower left of panel (a)).
Using
\begin{eqnarray}
q = \frac{2}{3}\frac{\rho_{\mathrm{av}}}{\rho_{\mathrm{av}} - \rho_{1/3}},
\label{eq-DWspacing2}
\end{eqnarray}
Eq.~\ref{eq-DWspacing1} yields the relation between $l$ and the {\it average} areal-density $\rho_{\mathrm{av}}$ of a striped DW phase:
\begin{eqnarray}
l^{-1} = \frac{\rho_{\mathrm{av}} - \rho_{1/3}}{b\rho_{1/3}}.
\label{eq-DWspacing3}
\end{eqnarray}
Equation~\ref{eq-DWspacing3} is plotted as the solid line in Fig.~\ref{Fig_suppl_DWdistance}(c); the labels indicate several integer-$q$ cases.

The critical density $\rho_{\mathrm{2}} =$ 7.22 $\pm$ 0.03\,nm$^{-2}$, which separates the $\alpha_{1}$ phase with variable DW spacing and the $\alpha_{2}$ phase with a fixed DW spacing ($l_{\mathrm{c}}$), lies close to the density for the six-row (integer-$q$ or $n$) configuration ($7.161$\,nm$^{-2}$).
However, given the $\pm2$\% absolute uncertainty of our density scale (see Sec.\,1), the true value of $\rho_{\mathrm{2}}$ could fall anywhere between 7.04 and 7.40\,nm$^{-2}$, a range that encompasses the $n =$ 5, 6, and 7 structures.
Thus, while the abrupt saturation of $\Delta S$ and other clear signatures at $\rho_{\mathrm{2}}$ (main text) provide stringent support for a fixed-row striped-DW state in the $\alpha_{2}$ phase---most likely $n = 6$---the absolute uncertainty of the density scale means that $n = 5$ or 7 cannot be excluded. \\
%
%
%
%
%
%
%
\section{Preliminary results for helium-4 on ZYX graphite}
\label{sec:preliminary}
\begin{figure}[h]
  \includegraphics[width=0.8\columnwidth]{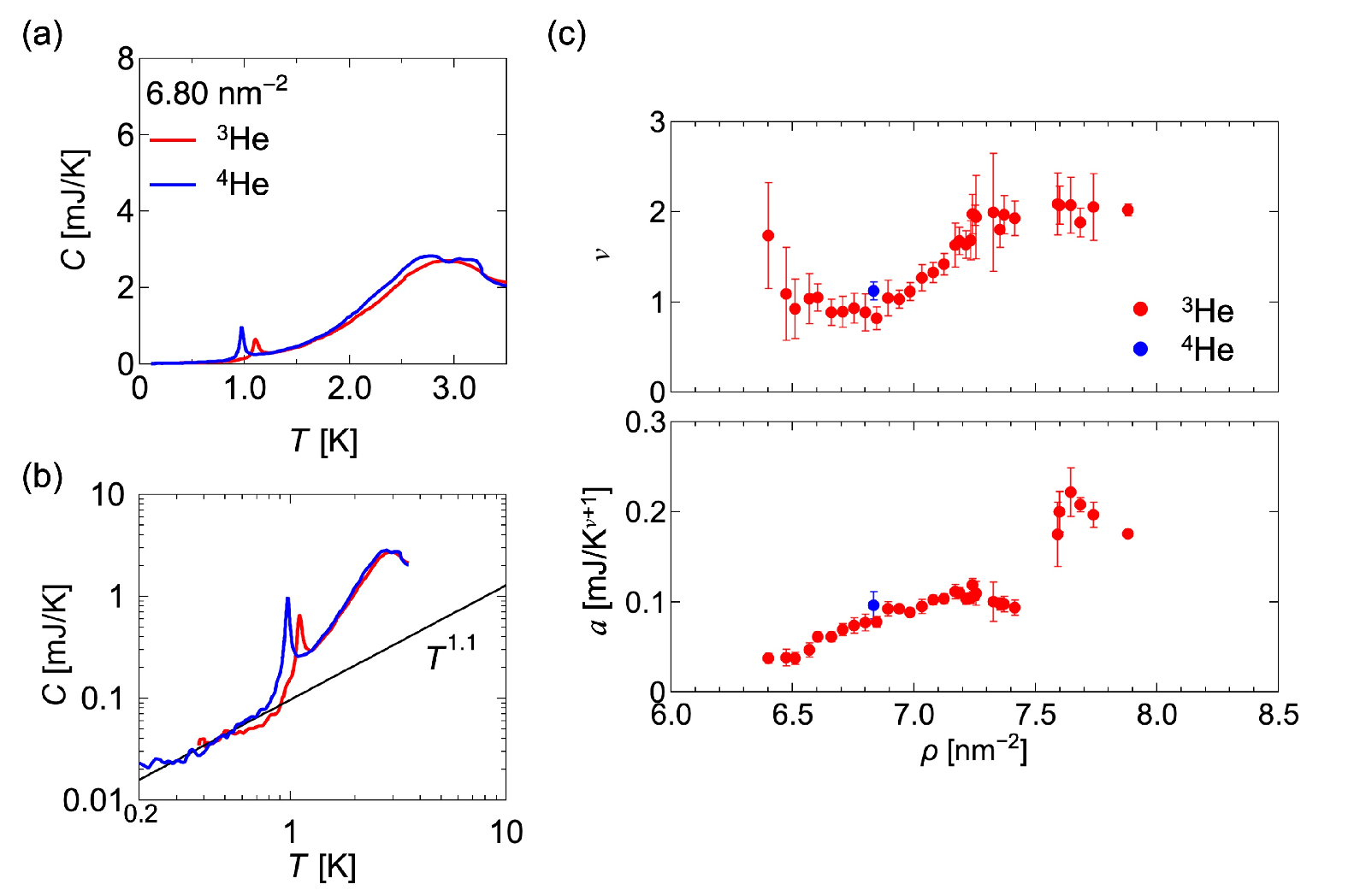}
  \caption{
(a) Heat capacities of $^3$He/ZYX (red curve) and $^4$He/ZYX (blue curve) at $\rho = 6.80\ \mathrm{nm}^{-2}$; (b) corresponding log--log plot.  
(c) Exponent $\nu$ and coefficient $a$ deduced from power-law fits $C = a T^{\nu}$ for the $^4$He/ZYX data shown in (a,b), compared with those for $^3$He/ZYX (same datasets as Fig.\,4).
}
  \label{Fig_suppl_3He4He}
\end{figure}

Figures~\ref{Fig_suppl_3He4He}(a) and (b) show the temperature dependences of the heat capacity for $^3$He/ZYX and $^4$He/ZYX at the same density $\rho = 6.80\ \mathrm{nm}^{-2}$ ($\alpha_1$ phase).  
Despite the different quantum statics, the two datasets exhibit remarkably similar behavior. 
In particular, the low-temperature $T$-linear heat capacity ($\nu \sim$ 1) and the coefficient $a$ are comparable in magnitude (Fig.~\ref{Fig_suppl_3He4He}(c)), indicating that the $T$-linear contribution reflects a phenomenon largely independent of quantum statistics.

The $^4$He system has a slightly lower $T_{\mathrm{peak}}$ for the 1\,K peak than the $^3$He system at the same density (Fig.~\ref{Fig_suppl_3He4He}(a)).
This accords with the known isotope effect in quantum solids and quantum liquid-crystals~\cite{nakamura2016possible}, and supports that the $\alpha_1$ phase possesses spatial order characteristic of a quantum DW phase (e.g., nematicity).
%
%
\clearpage